\documentclass[twocolumn,aps,prl,superscriptaddress]{revtex4}
\usepackage{graphicx}
\usepackage{amsmath}
\usepackage{bm}

\def\bea{\begin{eqnarray}}
\def\eea{\end{eqnarray}}
\def\be{\begin{equation}}
\def\ee{\end{equation}}

\DeclareMathOperator{\csch}{csch}

\begin{document}

\title{Thermometry via Light Shifts in Optical Lattices}

\author{M. McDonald}
\affiliation{Department of Physics, Columbia University, 538 West 120th Street, New York, NY 10027-5255, USA}
\author{B. H. McGuyer}
\affiliation{Department of Physics, Columbia University, 538 West 120th Street, New York, NY 10027-5255, USA}
\author{G. Z. Iwata}
\affiliation{Department of Physics, Columbia University, 538 West 120th Street, New York, NY 10027-5255, USA}
\author{T. Zelevinsky}
\email{tz@phys.columbia.edu}
\affiliation{Department of Physics, Columbia University, 538 West 120th Street, New York, NY 10027-5255, USA}

\begin{abstract}     
For atoms or molecules in optical lattices, conventional thermometry methods are often unsuitable due to low particle numbers or a lack of cycling transitions.  However, a differential spectroscopic light shift can map temperature onto the line shape with a low sensitivity to trap anharmonicity.  We study narrow molecular transitions to demonstrate precise frequency-based lattice thermometry, as well as carrier cooling.  This approach should be applicable down to nanokelvin temperatures.  We also discuss how the thermal light shift can affect the accuracy of optical lattice clocks.


\end{abstract}
\date{\today}
\maketitle

\newcommand{\w}{3.25in}

\newcommand{\Schematic}[1][\w]{
\begin{figure}[h]
\includegraphics*[trim = 0in 4.5in 2in 0.5in, clip, width=3.375in]{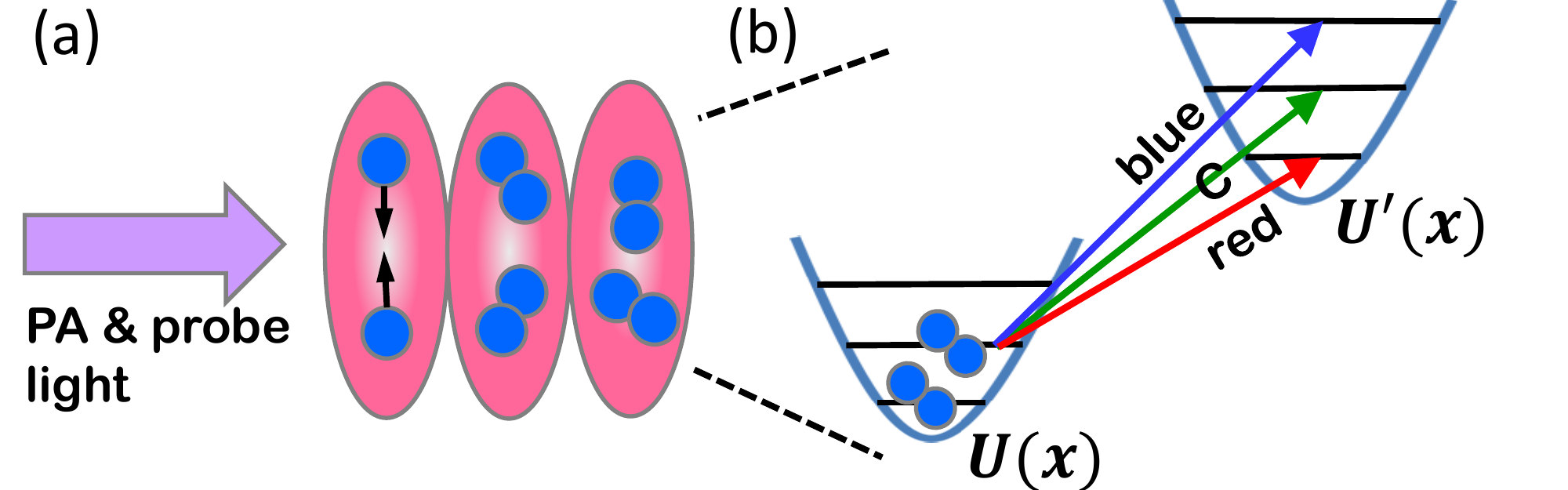}
\caption{(a) $^{88}$Sr atoms are trapped and cooled in a 1D optical lattice, and subsequently photoassociated on the narrow 689 nm intercombination line to create ultracold diatomic molecules.  The molecules are then probed along the lattice axis in the LD and RSB regimes.  (b) The carrier (C) and blue and red sideband (SB) transitions between long-lived electronic states in an approximately harmonic trap are indicated.}
\label{fig:Schematic}
\end{figure}
}

\newcommand{\LatticeSidebands}[1][\w]{
\begin{figure}[h]
\includegraphics*[trim = 0in 0in 0in 0in, clip, width=3.375in]{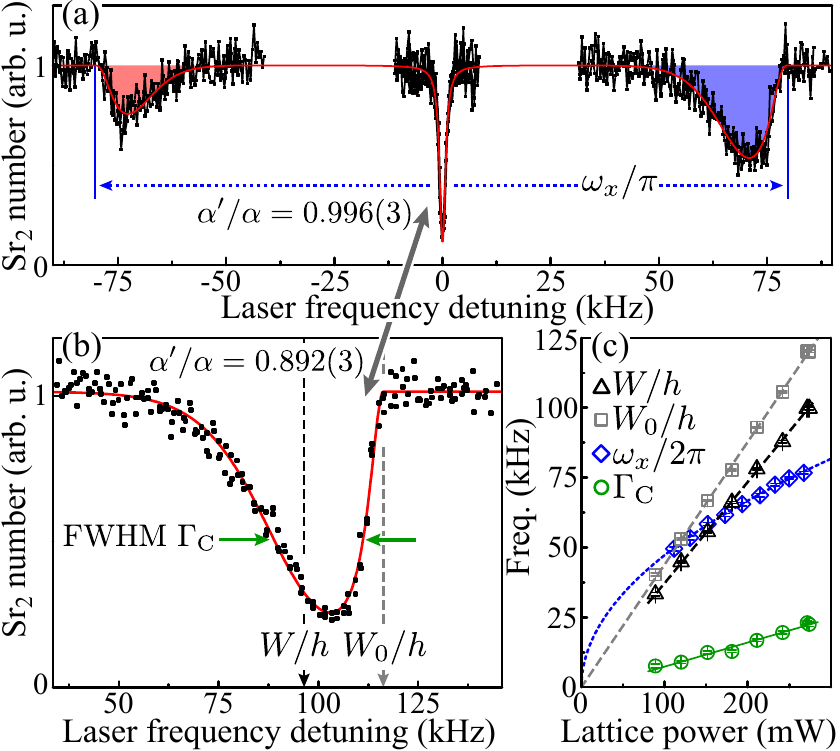}
\caption{(a) An optical spectrum of Sr$_2$ molecules in a state-insensitive lattice.  The central carrier transition and the first-order red and blue SBs are visible (the SB signals are enhanced via longer probing times).  The axial trap frequency $f_x\sim80$ kHz is found from the SB spacing \cite{Supplemental}, while $f_r\sim0.6$ kHz.  (b) The carrier line shape in a state-sensitive lattice, including light-induced shift and broadening. The average light shift, $W/h$, and the temperature-independent contribution to the light shift, $W_0/h$, are indicated. (The natural logarithm of the data was taken prior to fitting, to account for linear probe absorption.)  Zero detuning of the probe laser on the horizontal axes in (a) and (b) corresponds to zero lattice light shift. (c) The dependence of $W/h$, $W_0/h$, $f_x$, and $\Gamma_{\mathrm{C}}$ on the lattice light power.}
\label{fig:LatticeSidebands}
\end{figure}
}

\newcommand{\Temperature}[1][\w]{
\begin{figure}[h]
\includegraphics*[trim = 0in 0in 0in 0in, clip, width=3.375in]{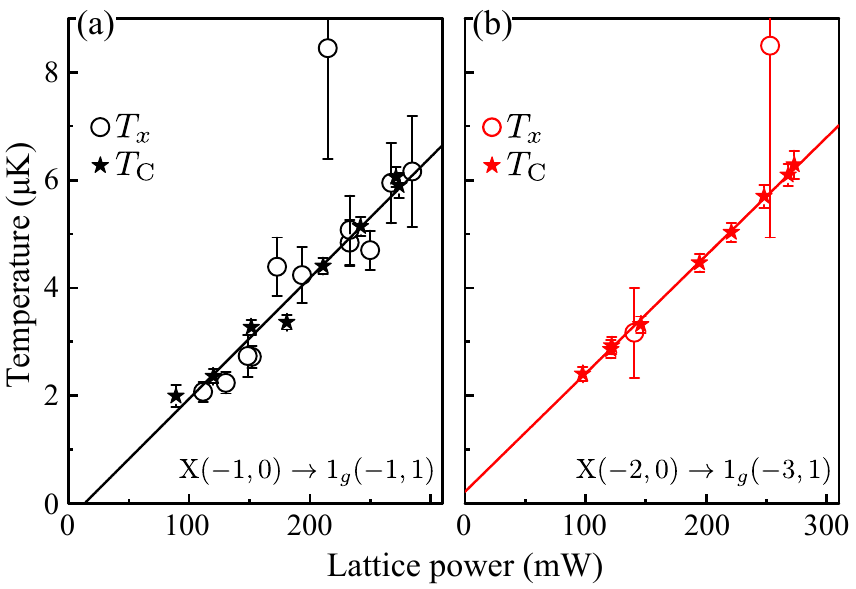}\hfill
\caption{Carrier thermometry of ultracold molecules (stars) is compared with an alternative technique that uses SB areas (circles), for (a) $v=-1$ and (b) $v=-2$ molecules.  The notation $(v,J)$ specifies the vibrational level and total angular momentum of the molecule.  The molecular potentials $X$ and $1_g$ \cite{ZelevinskyMcGuyerNPhys15_Sr2M1} are separated by an optical frequency.}
\label{fig:Temperature}
\end{figure}
}

\newcommand{\Systematics}[1][\w]{
\begin{figure}[h]
\includegraphics*[trim = 0in 0in 0in 0in, clip, width=3.375in]{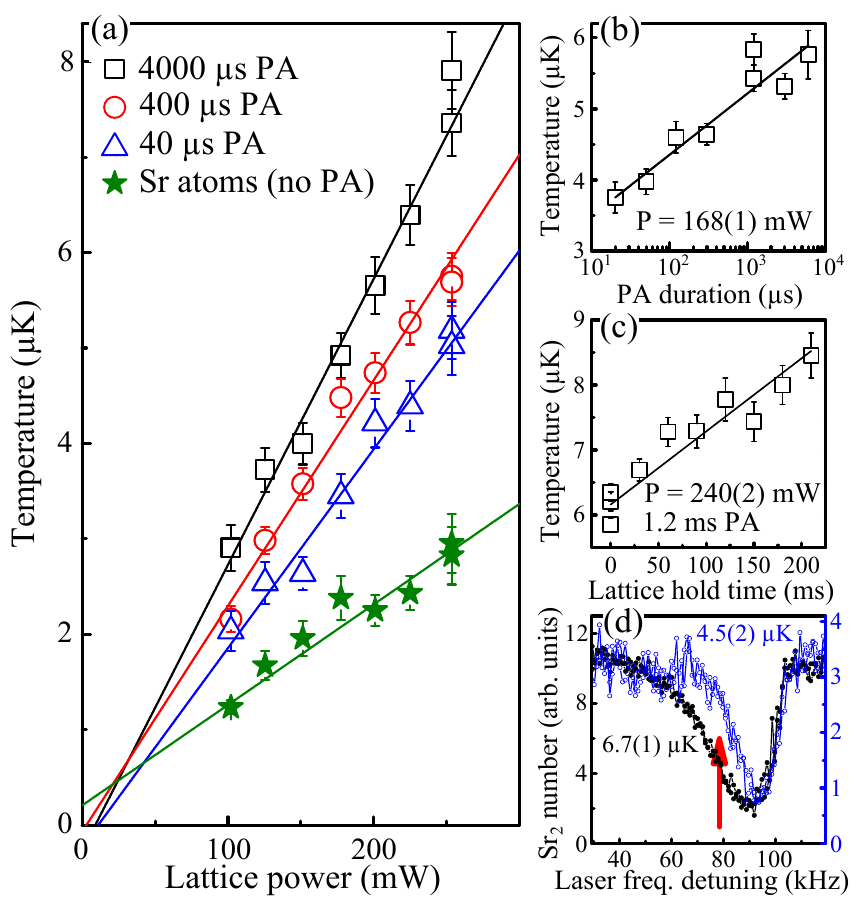}\hfill
\caption{(a) Molecule temperatures at various lattice light powers via carrier thermometry, along with the initial atom temperatures via TOF expansion imaging.  (b) Molecule temperature versus the PA pulse duration.  (c) Molecule temperature versus the lattice hold time.  (d) Carrier cooling of molecules in a weakly state-sensitive lattice.}
\label{fig:Systematics}
\end{figure}
}

Neutral atoms in tight optical traps have proven to be indispensable for time and frequency metrology \cite{KatoriNPhot11_LatticeClocks,LudlowHinkleyScience13_10to18YbComparison,LodewyckLeTargatNatureComm13_5SrClockNetwork,YeBloomNature14_10to18SrComparison,LisdatFalkeNJP14_SrLatticeClock} and studies of many-body quantum phenomena \cite{YeMartinScience13_SrClockManyBody}.  Recently, molecules have been similarly trapped and used for precision measurements of two-body and many-body physics \cite{ZelevinskyMcGuyerNPhys15_Sr2M1,ZelevinskyMcGuyerPRL13_Sr2ZeemanNonadiabatic,YeYanNature13_KRbLatticeSpinModel}.  These state-of-the-art experiments require ultracold temperatures to maximize control over the external degrees of freedom.  However, few reliable thermometry techniques exist aside from time-of-flight (TOF) expansion imaging \cite{MetcalfLettPRL88_SubDopplerCooling}.  This technique is unreliable with low particle numbers or a lack of cycling transitions, as is often the case for molecules \cite{YeWangPRA10_KRbAbsorpImaging}.  Moreover, existing methods often break down at ultralow temperatures in the nanokelvin regime, and new thermometry tools are needed, particulary those not relying on complex modeling \cite{DeMarcoMcKayRPP11_OpticalLatticeCooling}.

In this Letter, we show that if atoms or molecules can be trapped and probed in the Lamb-Dicke (LD) and resolved-sideband (RSB) regimes \cite{LeibfriedRMP03} of an optical lattice, the temperature $T$ can be determined from the spectrum of the carrier line (``C" in Fig. \ref{fig:Schematic}) with a precision that is roughly an order of magnitude higher than for conventional sideband (SB) based thermometry \cite{YeBlattPRA09_RabiSpect1DLattice}.  This temperature determination requires only the polarizability ratio $\alpha'/\alpha$ for the excited and ground states at the trap conditions, and the full-width-at-half-maximum (FWHM) $\Gamma_{\mathrm{C}}$ of the carrier line shape.  We show that
\be
T_{\mathrm{C}}\approx\frac{0.295\Gamma_{\mathrm{C}}}{|\sqrt{\alpha'/\alpha}-1|}\frac{h}{k_B},
\label{eq:TResult}
\ee
where $h=2\pi\hbar$ and $k_B$ are the Planck and Boltzmann constants, $\alpha',\alpha>0$, and $T_C$ refers to carrier-based thermometry.  A light shift measurement can directly yield $\alpha'/\alpha=1-2W_0/(M\lambda^2f_x^2)$ where $M$ is the particle mass, $\lambda$ is the lattice wavelength, $f_{x,r}\equiv\omega_{x,r}/(2\pi)$ are the trap frequencies along the axial and transverse directions defined relative to the probe beam, and $W_0$ is the temperature-independent light shift as described below.  Expression (\ref{eq:TResult}) is valid (i) for Boltzmann particle statistics in deep lattices, (ii) in the RSB regime $\Gamma_0<f_x$, where $\Gamma_0$ is the unbroadened carrier linewidth that would be measured in a "magic" trap with $\alpha'/\alpha=1$, (iii) if $\Gamma_0\ll\Gamma_C$, and (iv) if the transverse SBs are not excited, either by ensuring $\Gamma_0<f_r$ or by strictly controlling the probe-lattice alignment \cite{YeBlattPRA09_RabiSpect1DLattice}.  Importantly, $T_C$ is insensitive to leading-order trap anharmonicities.  This carrier thermometry should be valid at very low temperatures as long as $\alpha'/\alpha$ can be tuned sufficiently far from unity.  For example, if $\Gamma_0\sim1$ Hz and $\alpha'/\alpha\gtrsim1.2$, the subnanokelvin regime is accessible, provided there is a sufficient signal-to-noise ratio to detect the few particles in excited trap levels.

Carrier thermometry is applicable in the radio-frequency (rf), microwave, and optical regimes, which includes experiments with alkali and alkaline-earth metal atoms as well as simple molecules.
We apply the technique to accurately characterize the temperature of molecules created via photoassociation in an optical lattice.  We identify and quantify the heating mechanisms, and furthermore use the temperature dependence of the carrier for reducing the molecule temperature by a factor of $1.5$.

\Schematic
\LatticeSidebands
To demonstrate carrier thermometry, we create Sr$_2$ molecules via photoassociation (PA) \cite{JonesRMP06} from laser-cooled $^{88}$Sr atoms in a one-dimensional (1D) optical lattice \cite{ZelevinskyReinaudiPRL12_Sr2}, as illustrated in Fig. \ref{fig:Schematic}.  The lattice, which is present during laser cooling, is tunable by tens of nanometers around a central wavelength of 910 nm, and has a beam waist $w_0\sim25$ $\mu$m.  The probing is performed on an optical transition to a subradiant excited state of Sr$_2$ \cite{ZelevinskyMcGuyerNPhys15_Sr2M1}, with $\Gamma_0<200$ Hz limited by the laser linewidth.  Figure \ref{fig:LatticeSidebands}(a) shows an optical spectrum taken along the lattice axis, including the narrow carrier transition and the first-order axial SBs.  The trap frequency $f_x$ is found from the SB spacing with a state-insensitive lattice [$\alpha'/\alpha=0.996(3)$] \cite{ZelevinskyMcGuyerNPhys15_Sr2M1}.  As $\alpha'/\alpha$ is tuned via wavelength or polarization (the latter is possible via tensor shifts, and experimentally simpler), a differential light shift moves the line center, and an asymmetric $T$-dependent line broadening develops.  The carrier spectrum in Fig. \ref{fig:LatticeSidebands}(b) corresponds to $\alpha'/\alpha=0.892(3)$.  The line is thermally broadened, with the FWHM $\Gamma_{\mathrm{C}}$ yielding $T_{\mathrm{C}}$ in Eq. (\ref{eq:TResult}).

\Temperature
The measured temperatures $T_{\mathrm{C}}$ are plotted in Fig. \ref{fig:Temperature} for various optical lattice light powers, for Sr$_2$ molecules in the two least-bound vibrational levels of the electronic ground state $X$ ($v=-1$ and $v=-2$).  Also shown are axial temperatures independently determined from the ratios of the blue to red SB areas \cite{YeBlattPRA09_RabiSpect1DLattice}
\be
T_x\approx\frac{\hbar\omega_x}{k_B}\left(\ln\frac{A_{\mathrm{blue}}}{A_{\mathrm{red}}}\right)^{-1}.
\label{eq:TRatio}
\ee
Equation (\ref{eq:TRatio}) holds for Boltzmann statistics in deep lattices (in our experiment, trap depth $U_0\sim10k_BT$).
While the methods are in close agreement, carrier thermometry benefits from the larger signal-to-noise ratio for the carrier relative to the SBs.  Its uncertainties are smaller by almost an order of magnitude, reaching the percent level.  The smaller number of $v=-2$ molecules in Fig. \ref{fig:Temperature}(b) hampers the SB but not the carrier measurements.  Moreover, at colder temperatures than are reached here, the carrier method is expected to be superior due to the large relative uncertainty of determining the area of a vanishingly small red SB.

\Systematics
We apply carrier thermometry to systematically characterize the temperature of photoassociated molecules.  Since PA is the optimal method to create a variety of dimers and can yield molecules in the absolute ground state \cite{MoszynskiSkomorowskiPRA12_Sr2Formation,DeMilleBruzewiczNJP14_GroundStateRbCsViaPA}, any heating that occurs during PA must be understood  and suppressed if such molecules are to reach a high phase-space density.  PA into an electronically excited molecular level is usually followed by a spontaneous or stimulated decay to the electronic ground state.  From kinetic considerations, the temperature of homonuclear dimers created in this way is expected to be nearly the same as for the precursor atoms \cite{JulienneCiuryloPRA04_IntercombinationPA}.  However, for typical conditions we observe heating by roughly a factor of 2, as shown in Fig. \ref{fig:Systematics}(a).  This heating is caused by incoherent photon scattering and can be partially suppressed by using shorter PA pulse durations, as shown in Figs. \ref{fig:Systematics}(a) and \ref{fig:Systematics}(b).  Heating of the molecules by the optical lattice [Fig. \ref{fig:Systematics}(c)] \cite{MeschedeAltPRA03_SingleAtomsInLattice} is also present, but occurs at a much slower rate.

While SB cooling is an established technique for reducing axial temperatures \cite{WinelandDiedrichPRL89_SidebandCooling}, the method described here suggests a "carrier cooling" procedure that can reduce the 3D temperature of a cold gas.  Figure \ref{fig:Systematics}(d) illustrates the reduction of the molecule temperature by a factor of $1.5$ via weakly exciting the hotter molecules in the tail of the line shape.  In the case of open transitions such as for Sr$_2$, the lowering of $T$ is due to energy selection
\cite{DavidsonKaplanJOB05_RFSpectroscInOptTraps} and is not accompanied by a phase-space density increase.

The experiments in this work utilize a narrow optical transition, but carrier thermometry can be performed in any frequency regime.  We have confirmed the results for a two-photon $1.3$ GHz vibrational transition in ground-state Sr$_2$.  In microwave and rf regimes, the LD and RSB conditions for carrier thermometry may be satisfied in optical traps not utilizing a lattice configuration.

To obtain Eq. (\ref{eq:TResult}), we consider the case of a 1D lattice to reflect our setup as well as that of many lattice-clock type experiments, but note that the method is general and extends to 2D and 3D lattices.  Near the center of a lattice well, the potential energy is nearly harmonic
\begin{align}
U({\bf r})\approx\frac{1}{2} M \omega_x^2 x^2 + \frac{1}{2} M \omega_r^2 (y^2 + z^2) - U_0,
\label{Eq:Ur}
\end{align}
where
\begin{align}
\omega_x = ({2 \pi}/{\lambda}) \sqrt{{2 U_0}/{M}}
\;\; \text{and} \;\;
\omega_r = ({2}/{w_0}) \sqrt{{U_0}/{M}},
\label{Eq:WxWr}
\end{align}
$U_0=4\alpha P/(\pi w_0^2c\epsilon_0)$, $P$ is the lattice light power,
$c$ is the speed of light, and $\epsilon_0$ is the permittivity of vacuum.  If $\alpha'/\alpha\neq1$, the potential for the final internal state differs from that for the initial state, $U'({\bf r})\neq U({\bf r})$.  The light shift (or differential ac Stark shift) of a carrier transition by the lattice is the difference in final and initial expectations
\begin{align}
W &= \langle H'\rangle - \langle H\rangle.
\label{Eq:W1}
\end{align}
Using Eq. (\ref{Eq:Ur}), the energy expectation values are of the form
\begin{align}
\langle H\rangle= \hbar \omega_x \langle n_x +1/2\rangle + \hbar \omega_r \langle n_r+ 1 \rangle  - U_0,
\label{Eq:Hi}
\end{align}
where $H$ is the single-particle Hamiltonian, $n_i$ are harmonic oscillator occupation numbers, and $n_r\equiv n_y+n_z$.  Evaluating the net light shift (\ref{Eq:W1}) using Eq. (\ref{Eq:Hi}), under the assumption that SBs are not excited at an appreciable rate such that $n_i'=n_i$ \cite{KimJKPS07_TrapInducedBroadening}, we find
\begin{align}
W =& W_0 + W_x + W_r \label{Eq:Wparts}=\left( 1 - \frac{\alpha'}{\alpha} \right) U_0 + \left( \sqrt{\frac{\alpha'}{\alpha}} - 1 \right)\nonumber \\
	&\times\left( \frac{\hbar \omega_x}{2} \text{coth}\left[ \frac{ \hbar \omega_x }{ 2 k_B T } \right]+ \hbar \omega_r \, \text{coth}\left[ \frac{ \hbar \omega_r }{ 2 k_B T } \right] \right).
\end{align}
Note that if equipartition is valid ($\hbar\omega_{x,r}\ll k_BT$), Eq. (\ref{Eq:Wparts}) simplifies to
\begin{align}
W &\approx \left( 1 - {\alpha'}/{\alpha} \right) U_0 + 3 \left( \sqrt{{\alpha'}/{\alpha}} - 1 \right) k_B T,
\label{Eq:WLinear}
\end{align}
highlighting the nonthermal and thermal contributions to $W$.
The carrier line shape described below permits a clean extraction of $W_0=(1-\alpha'/\alpha)U_0$, and therefore of $\alpha'/\alpha$, if the trap depth or axial trap frequency (\ref{Eq:WxWr}) is known.  The measured shifts $W$ and $W_0$ are marked in Fig. \ref{fig:LatticeSidebands}(b) and plotted versus the lattice light power in Fig. \ref{fig:LatticeSidebands}(c).

The $T$-dependent light shifts $W_x$ and $W_r$ in Eq. (\ref{Eq:Wparts}) cause asymmetric line broadening \cite{KatoriTakamotoPRL03_SrClockSpectrosc}.  The carrier transition from the trap state $|n_x\; n_r\rangle$ experiences a differential light shift
\begin{align}
\delta E =	\delta E_x + \delta E_r,
\label{Eq:dE1}
\end{align}
where the axial and radial contributions are
\begin{align}
\delta E_x &= (\sqrt{\alpha'/\alpha}-1)\hbar\omega_x(n_x + 1/2), \label{dEx} \\
\delta E_r &= (\sqrt{\alpha'/\alpha}-1)\hbar\omega_r(n_r+ 1), \label{dEr}
\end{align}
and $\langle \delta E_{x,r} \rangle=W_{x,r}$.  The Boltzmann probability distribution for the discrete variable $\delta E_x$ is
\begin{align}
p_x(\delta E_x) =\frac{1}{Z_x}e^{-u(\delta E_x)}.
\label{Eq:pEx}
\end{align}
The partition function is $Z_i=\frac{1}{2}\csch\frac{\hbar\omega_i}{2k_BT}$, and the dimensionless function
\be
u(\delta E_i)=\frac{ \delta E_i}{ k_B T (\sqrt{\alpha' / \alpha} - 1) } \geq 0
\label{Eq:u}
\ee
parametrizes the Boltzmann exponent with a discrete step size of $\Delta_i=\hbar\omega_i/(k_BT)$ for $i=x,r$.
Similarly, for the radial shift,
\be
p_r(\delta E_r)=\frac{1}{Z_r^2}\frac{1}{\Delta_r}u(\delta E_r)e^{-u(\delta E_r)}.
\label{Eq:pEr}
\ee
The discrete probability for the energy $\delta E$ is then the convolution
\begin{align}
p&(\delta E) = \sum_{\{n_x,n_r\}_{\delta E}} p_x(n_x) p_r(n_r)
\label{eq:pDiscrete}
\end{align}
over the pairs of $n_x$ and $n_r$ satisfying $\delta E(n_x,n_r) = \delta E$.

If $\Delta_x\lesssim1$, which is the case here, the discrete expression (\ref{eq:pDiscrete}) may be simplified in the continuum limit of $\Delta_i\rightarrow0$.  Noting that the probability density
\begin{align}
\overline{p_i}[u(\delta E_i)]= \lim_{\Delta_i \rightarrow 0} p_i(\delta E_i)/\Delta_i,
\end{align}
we obtain $\overline{p}_x=e^{-u(\delta E_x)}$ and $\overline{p}_r=u(\delta E_r)e^{-u(\delta E_r)}$.
The probability $p(\delta E)$ then reduces to a gamma distribution
\be
\overline{p}[u(\delta E)] = \int_0^\infty \overline{p}_r(u - u_x) \overline{p}_x(u_x) d u_x= \frac{1}{2}u^2 e^{-u}.
\label{Eq:Lineshape}
\ee
The probability density (\ref{Eq:Lineshape}) directly yields the spectroscopic line shape since carrier transition rates are nearly independent of $n_i$ (this assumption may need to be modified if $\alpha'/\alpha$ is far from unity).
The line shape has the form of a Boltzmann distribution in a 3D harmonic trap \cite{Metcalf}, as can be expected from (\ref{Eq:dE1})-(\ref{dEr}).
In a special case where only the ground axial state is occupied ($\Delta_x\gtrsim5$), the line shape (\ref{Eq:Lineshape}) is replaced by the 2D Boltzmann result $\overline{p}[u(\delta E)] =ue^{-u}$.  We have assumed $T_x=T_r\equiv T$ \cite{YeBlattPRA09_RabiSpect1DLattice}, but the analysis can be adapted to other situations, including non-Boltzmann distributions.

The FWHM of the function (\ref{Eq:Lineshape}) is nearly 3.395, which together with (\ref{Eq:u}) yields Eq. (\ref{eq:TResult}).  Figure \ref{fig:LatticeSidebands}(b) shows a fit of line shape (\ref{Eq:Lineshape}) to a carrier spectrum, with its FWHM $\Gamma_{\mathrm{C}}$ directly yielding temperatures in Figs. \ref{fig:Temperature} and \ref{fig:Systematics}.  Note that the width of the carrier is much smaller than its light shift; hence, it was necessary to stabilize $P$.  This was done with minimal error from interference of the forward and retroreflected lattice beams by using a pellicle beam splitter to sample the forward beam produced by an optical fiber with an angled output face.

It is a property of harmonic oscillator eigenstates that the dominant anharmonic corrections (proportional to $x^4$, $x^2r^2$, $r^4$) shift their energy by an amount that is independent of trap depth \cite{Supplemental}.  Hence Eqs. (\ref{Eq:Lineshape}) and (\ref{eq:TResult}) are largely unaffected by anharmonic corrections.  However, these corrections allow a determination of temperature from the shape of first-order axial SBs in a 1D lattice, using spectra with $\alpha'/\alpha=1$ as in Fig. \ref{fig:LatticeSidebands}(a) \cite{YeBlattPRA09_RabiSpect1DLattice}.  As for carriers, this approach yields 3D temperatures [versus 1D temperatures for Eq. (\ref{eq:TRatio})], where sensitivity to $T$ now arises from the anharmonicity of the lattice trap.  Introducing leading-order corrections to the harmonic approximation of a sinusoidal potential, and adapting the approach used to derive the carrier line shape \cite{Supplemental}, we find the temperature in a state-insensitive lattice
\be
T_{\mathrm{SB}}\approx0.484M\lambda^2f_x\Gamma_{\mathrm{SB}}/k_B,
\label{eq:SBWidth}
\ee
where $\Gamma_{\mathrm{SB}}$ is the FWHM of the SB line shape given by $\overline{p}[v(\delta E)] =(1/6)v^3 e^{-v}$, and $v$ is a function similar to $u$ \cite{Supplemental}.  Using this approach on the data of Fig. \ref{fig:Temperature}(a) yields $T_{\mathrm{SB}}$ that is too high by $\sim1$ $\mu$K compared to $T_{\mathrm{C}}$ and $T_x$.  There are several reasons for Eq. (\ref{eq:SBWidth}) to be less reliable than Eqs. (\ref{eq:TResult}) and (\ref{eq:TRatio}).  First, SBs are more sensitive than the carrier to distortion by other broadening mechanisms, since there is no tunability of $\Gamma_{\mathrm{SB}}$, unlike for $\Gamma_C$.  Additionally, any axial displacement from the Gaussian lattice beam waist produces new leading-order anharmonic corrections to Eqs. (\ref{Eq:Ur}) and (\ref{eq:SBWidth}) that could strongly affect the SB result.  More generally, $T_{\mathrm{C}}$ in Eq. (\ref{eq:TResult}) depends only on the polarizability ratio $\alpha'/\alpha$, a fundamental property of the atom or molecule that can be measured with a high accuracy.  In contrast, $T_{\mathrm{SB}}$ depends on $f_x$, which varies slightly across the sample.

Note that the thermal contribution $(W-W_0)/h$ in Fig. \ref{fig:LatticeSidebands}(c) and Eqs. (\ref{Eq:Wparts}) and (\ref{Eq:WLinear}) is a significant fraction of the lattice light shift.  This could affect optical lattice clocks \cite{LudlowHinkleyScience13_10to18YbComparison,LodewyckLeTargatNatureComm13_5SrClockNetwork,YeBloomNature14_10to18SrComparison,LisdatFalkeNJP14_SrLatticeClock} if the atom temperature versus trap depth [Fig. \ref{fig:Systematics}(a), stars] does not linearly extrapolate to exactly zero at $P=0$ \cite{PhillipsGatzkePRA97_TemperatureOfAtomsInLattices,WeissWinotoPRA99_CoolingInLattices}.  For example, if it extrapolates to just $\pm0.1$ $\mu$K, then for $\alpha'/\alpha\sim1\pm3\times10^{-7}$ as in Ref. \cite{YeBloomNature14_10to18SrComparison}, the residual thermal line pulling from Eq. (\ref{Eq:WLinear}) is $\sim3\times10^{-18}$, comparable to the total uncertainty budget.  Furthermore, for typical clock experiment conditions, the full clock shift (\ref{Eq:Wparts}) must be used, where $W$ is not linear in $T$.  This nonlinearity leads to an effective offset $|T(P=0)|\sim0.1$ $\mu$K in an experiment with our parameters, again leading to line pulling.  Counterintuitively, if the temperature is kept fixed at all lattice depths, this thermal pulling is even more problematic.  In general, the variation of temperature with lattice depth depends on the cooling and trapping procedure.

In conclusion, we have shown that narrow spectral lines of atoms or molecules tightly trapped in optical lattices allow highly precise 3D temperature determinations, and are not limited by low particle numbers, a lack of cycling transitions, or ultralow temperatures.  The method is purely frequency based, requiring only measurements of the carrier linewidth, light shift, and axial trap frequency, and is mostly immune to trap anharmonicities.  We experimentally demonstrate complete control over molecular external and internal degrees of freedom in the LD and RSB regimes of a weakly state-sensitive optical lattice, use carrier cooling to reduce the temperature of the ultracold molecules, and identify the significant heating processes of photoassociated molecules.  Furthermore, the result (\ref{eq:TResult}) can be inverted to accurately predict light-shift-induced thermal dephasing.

We thank M. G. Tarallo, A. T. Grier, and S. Rolston for discussions, and acknowledge ONR Grant No. N00014-14-1-0802, NIST Grant No. 60NANB13D163, and ARO Grant No. W911NF-09-1-0504 for partial support of this work.  M. M. and G. Z. I. acknowledge NSF IGERT Grant No. DGE-1069260.

\section{Supplemental Material}

\subsection{Anharmonicity for carriers}
To address the effects of anharmonicity, we consider the model potential
\begin{align}	\label{Umodel1D}
U({\bf r}) \approx - U_0 \, e^{-2 \, (y^2 + z^2)/ w_0^2} \, \cos^2 (2 \pi x / \lambda)
\end{align}
for a 1D optical lattice, which is a good approximation near the trap center.
This potential introduces three leading-order anharmonic corrections to (3), which are the quartic potentials
\begin{align}		
V_{xx}({\bf r}) &= - \left(M \omega_x^2 \, x^2 / 2 \right)^2 / ( 3 U_0 ) 						\label{Vxx} \\
V_{xr}({\bf r}) &= - \left(M \omega_x^2 \, x^2 / 2 \right) \left( M \omega_r^2 \, r^2/2 \right) / U_0  	\label{Vxr} \\
V_{rr}({\bf r}) &= - \left(M \omega_r^2 \, r^2 / 2 \right)^2  / ( 2 U_0 ) 						\label{Vrr}
\end{align}
for the initial and likewise for the final lattice, where $r^2 \equiv y^2 + z^2$.
Far from the axial trap center, a finite Rayleigh length introduces additional (e.g., cubic) leading-order corrections.

For a transition between a pair of known trap states in the initial and final lattices we may approximate the light shift of each $V_{ij}({\bf r})$ by its first-order perturbation,
\begin{align}	\label{dEij}
\delta E_{ij} \approx \langle n_x' n_y' n_z' | V_{ij}' | n_x' n_y' n_z' \rangle - \langle n_x n_y n_z | V_{ij} | n_x n_y n_z \rangle,
\end{align}
where primes denote final-lattice values.
These shifts introduce the corrections
\begin{align} 	\label{Wij}
W_{ij} = \langle \delta E_{ij} \rangle
\end{align}
to the total light shift $W$ of (7),
where the brackets denote a thermal average over the allowed pairs of initial and final trap states.
For axial sideband (SB) transitions, the effects of $n_x$-dependent excitation rates must be included in this average, as described in the next section.
In the Lamb-Dicke (LD) and resolved-sideband regimes with suppressed transverse SB transitions, the trap state pairs satisfy
\begin{align} 	\label{ortho}
n_x' = n_x + D, ~n_y' = n_y, ~\text{and}~ n_z' = n_z,
\end{align}
where the integer $D$ is introduced to distinguish between axial carrier ($D = 0$) and first-order axial SB transitions ($D=\pm1$).

Surprisingly, carrier transitions are nearly unchanged by the leading-order corrections (\ref{Vxx}--\ref{Vrr}), because the first-order light shifts (\ref{dEij},\ref{Wij}) are zero:
\begin{align}
\delta E_{ij} = W_{ij} = 0 \quad \text{if} \quad D = 0.
\end{align}
Before evaluating these quantities explicitly in the next section, we can explain this general result as follows.
First, note that any form for $U({\bf r})$ must be proportional to the polarizability $\alpha$.
Thus, any anharmonic corrections to (3), such as (\ref{Vxx}--\ref{Vrr}), must also be proportional to $\alpha$.
Next, note that the expectations $\langle n_x | x^2 | n_x \rangle \propto 1/\sqrt{\alpha}$ and $\langle n_x | x^4 | n_x \rangle \propto 1/\alpha$ for harmonic oscillator states.
The matrix elements in (\ref{dEij}) for the $V_{ij}$ of (\ref{Vxx}-\ref{Vrr}) are therefore independent of $\alpha'$ and $\alpha$, respectively,
and must be equal, thus producing no differential shift.
This general insensitivity of the carrier light shift to quartic anharmonicities also applies to the model potentials
$- U_0 \, e^{-2 \, (z / w_0)^2} \, \cos^2 (2 \pi x / \lambda) \cos^2 (2 \pi y / \lambda)$  and
$- U_0 \cos^2 (2 \pi x / \lambda) \cos^2 (2 \pi y / \lambda) \cos^2 (2 \pi z / \lambda)$
for 2D and 3D optical lattices.

\subsection{Axial sideband transitions}

For axial SB transitions with $D \neq 0$, the total light shift $W$ of (7) due to the harmonic potential (3) must be modified as follows.
First, there is an ``axial-SB shift'' from the final lattice,
\begin{align}		\label{Ws}
W_s &= \hbar \omega_x' \, D,
\end{align}
which must be added as a fourth part to $W$.
Next, if $D < 0$, the populations of the initial lattice with $n_x < |D|$ will not participate in the transition, so the expectation $\langle n_x + 1/2 \rangle$ must be computed accordingly.  This asymmetry also leads to the relation (2) between temperature and the ratio of SB areas.

Additionally, for SB transitions the excitation rates depend on $n_x$.
The expectation $\langle n_x + 1/2 \rangle$ is no longer solely thermal, but must account for this inhomogeneous excitation by weighting each value of $n_x$ with the square of its Rabi frequency for the transition,
\begin{align}	\label{Rabi}
\Omega(n_x, D)^2 \propto
|\langle n_x' | e^{i k x} | n_x \rangle|^2 \approx
\begin{cases}
1 			& D = 0 \\
\eta^2 \, n_x 		& D = -1 \\
\eta^2 \, (n_x + 1) 	& D = +1,
\end{cases}
\end{align}
where the LD parameter $\eta = k \sqrt{\hbar/(2 M \omega_x)}$ and the axial wavenumber $k = 2 \pi/\lambda$.
As before, we assume the trap states are approximately orthonormal, $\langle n_x' | n_x \rangle \approx \delta_{n_x',n_x}$, which may need to be modified if $\alpha'/\alpha$ is far from unity.
After normalizing the probabilities for each $n_x$, the weighted expectations are
\begin{align} 	\label{nxWithD}
\langle n_x + \frac{1}{2} \rangle =
\begin{cases}
\text{coth}\left[ \Delta_x/2 \right]/2	& D = 0 \\
\text{coth}\left[ \Delta_x/2 \right]+1/2	& D = -1 \\
\text{coth}\left[ \Delta_x/2 \right]-1/2 	& D = +1
\end{cases}
\end{align}
where as before $\Delta_x = \hbar \omega_x/(k_B T)$.

Hence, although the form of $W_x = \langle \delta E_x \rangle$ given by (10),
\begin{align}		\label{Wx}
W_x = \left( \sqrt{\alpha'/\alpha} - 1 \right) \hbar \omega_x \langle n_x + 1/2 \rangle,
\end{align}
will be unchanged for SBs, the value of $W_x$ will depend on $D$ following (\ref{nxWithD}).
Note that the form and value of $W_r = \langle \delta E_r \rangle$ given by (11),
\begin{align}		\label{Wr}
W_r = \left( \sqrt{\alpha'/\alpha} - 1 \right) \hbar \omega_r \langle n_r + 1 \rangle,
\end{align}
is the same for SBs as for carriers.

The anharmonic corrections (\ref{Vxx}--\ref{Vrr}) are important for SBs, unlike carriers, especially for state-insensitive `magic' traps with $\alpha'/\alpha=1$.
The contributions (\ref{Wij}) to the shift $W$ from these corrections are
\begin{align}	
W_{xx} &= - \frac{W_s}{4 U_0} \left( \sqrt{\frac{\alpha}{\alpha'}} \, \hbar \omega_x \langle n_x + 1/2 \rangle + \frac{W_s \, \alpha}{2 \, \alpha'} \, \right)  	\label{Wxx} \\
W_{xr} &= - \frac{W_s}{4 U_0} \sqrt{\frac{\alpha}{\alpha'}} \, \hbar \omega_r \langle n_r + 1 \rangle		\label{Wxr} \\ 
W_{rr} &= 0  	\label{Wrr}
\end{align}
for both carrier and SB transitions, as derived below.
Importantly, note that all these contributions are zero for carriers as argued above, since $W_s = 0$ if $D = 0$.

The expression (\ref{Wxx}) for $W_{xx}$ follows from the expectation $\langle n_x | (M \omega_x^2 x^2 / 2)^2 | n_x \rangle = 3 (\hbar \omega_x)^2 (2n_x^2 + 2n_x + 1)/16$ \cite{LandauQM}, which gives the matrix elements
\begin{align}	\label{VxxElement}
\langle n_x n_y n_z | V_{xx} | n_x n_y n_z \rangle = - \frac{(\hbar \omega_x)^2}{16 U_0} ( 2n_x^2 + 2n_x + 1 ),
\end{align}
and from noting that $(\omega_x')^2/U_0' = \omega_x^2/U_0$ and $2(n_x')^2 + 2 (n_x') + 1 = 2 n_x^2 + 2 n_x + 1 + 4 D (n_x + 1/2) + 2 D^2$.
The expression for $W_{xr}$ follows from expectations of the form $\langle n_x | (M \omega_x^2 x^2/2) | n_x \rangle = \hbar \omega_x (n_x + 1/2)/2$,
which give the matrix elements
\begin{align}	\label{VxrElement}
\langle n_x n_y n_z | V_{xr} | n_x n_y n_z \rangle = - \frac{ \hbar \omega_x \hbar \omega_r }{4 U_0} ( n_x + 1/2 ) ( n_r + 1 ),  
\end{align}
and from noting that $\omega_x' \omega_r' / U_0' = \omega_x \omega_r / U_0$.
The shift $W_{rr}$ of (\ref{Wrr}) is then zero because the condition (\ref{ortho}) includes only radial carrier transitions.
That is, $V_{rr}$ of (\ref{Vrr}) contributes no shift for the same reasons that $W_{xx} = W_{xr} = 0$ if $D=0$.

To demonstrate the effects of anharmonicity on the lineshape of SB transitions, let us treat the case of a magic lattice with $\alpha'/\alpha=1$ where there is no thermal broadening of the carrier.
In this case, broadening comes only from the thermal distribution of the anharmonic shifts $\delta E_{xx}$ and $\delta E_{xr}$.
Using (\ref{VxrElement}) with (\ref{dEij}), we find
\begin{align} \label{dExr}
\delta E_{xr}(n_r) = -D (n_r + 1) \, \hbar \omega_r \, \hbar \omega_x / (4 U_0 ).
\end{align}
Similarly, using (\ref{VxxElement}) with (\ref{dEij}) and (\ref{Ws}),
\begin{align} 	\label{dExx}
\delta E_{xx}(n_x) 
	&= - \left[ 2 D (n_x + 1/2) (\hbar \omega_x)^2 + W_s^2 \right]/(8 U_0)  \nonumber \\ 
	&\approx -D (n_x + 1/2) (\hbar \omega_x)^2 / (4 U_0),
\end{align}
where the second line follows from neglecting a constant offset (half the lattice-photon recoil energy) that contributes no broadening.  Note that (\ref{Wxx},\ref{Wxr}) are related to (\ref{dExx},\ref{dExr}) via (\ref{Wij}) with $\alpha'/\alpha=1$.

Together, the shifts (\ref{dExr},\ref{dExx}) lead to similar lineshapes as derived for carrier transitions.
As before, we introduce a function to replace Boltzmann exponents,
\begin{align}	\label{vFunction}
v(\delta E) = -\delta E / [k_B T \, \hbar \omega_x \, D / (4 U_0)] \geq 0.
\end{align}
The discrete step size of $v(\delta E_{xx})$ is $\Delta_x = \hbar \omega_x/(k_B T)$ and of $v(\delta E_{xr})$ is $\Delta_r = \hbar \omega_r / (k_B T)$.
Since the probability distribution for $n_r$ is unchanged, the probability for the discrete variable $\delta E_{xr}$ follows from the $p_r$ of (14),
\begin{align}
p_{xr}(\delta E_{xr}) =  \frac{1}{Z_r^2 \Delta_r} \, v(\delta E_{xr}) \, e^{-v(\delta E_{xr})}.
\end{align}
Likewise, for $D \geq 0$ the probability $p_x$ of (12) for $n_x$ is unchanged.
However, we now need to account for inhomogeneous excitation, so the probability for the discrete variable $\delta E_{xx}$ is
\begin{align}
p_{xx}(n_x) \propto \Omega(n_x, D)^2 \, p_x(n_x).
\end{align}
For $D = 1$, using (\ref{Rabi}) and normalizing, this evaluates to
\begin{align}
p_{xx}(\delta E_{xx}) = \frac{v(\delta E_{xx}) + \Delta_x/2}{Z_x \Delta_x (1 + e^{-\Delta_x/2} Z_x)} \, e^{-v(\delta E_{xx})}.
\end{align}
Likewise, for $D = -1$ where only $n_x \geq 1$ participate,
\begin{align}
p_{xx}(\delta E_{xx}) = \frac{v(\delta E_{xx}) - \Delta_x/2}{Z_x^2 \Delta_x } \, e^{-v(\delta E_{xx})+\Delta_x/2}.
\end{align}
In the continuum limit, these probabilities simplify to 
\begin{align}
\overline{p}_{xi}[v(\delta E_{xi})] = \lim_{\Delta_i \rightarrow 0} \frac{p_{xi}(\delta E_{xi})}{\Delta_i} = v \, e^{-v}
\end{align}
for both $i = x,r$ and $D = \pm 1$.

Following (15), the distribution for the total shift $\delta E(n_x, n_r) = \delta E_{xx}(n_x) + \delta E_{xr}(n_r)$ is the convolution
\begin{align}	\label{pdEsideband}
p(\delta E) = \sum_{\{n_x,n_r\}_{\delta E}} p_{xx}(n_x) p_{xr}(n_r),
\end{align}
over the pairs of $n_x$ and $n_r$ satisfying $\delta E(n_x, n_r) = \delta E$.
In the continuum limit
this reduces to a Gamma distribution similar to (17),
\begin{align}	\label{pv}
\overline{p}[v(\delta E)] = \lim_{\Delta_x, \Delta_r \rightarrow 0} \, \frac{p(\delta E)}{\Delta_x \Delta_r} = \frac{1}{6} \, v^3 \, e^{-v},
\end{align}
for both $D = \pm 1$ SBs.
As expected and demonstrated in Fig.~2(a), the sharp edge of this lineshape is furthest from the carrier.
To extract axial trap frequencies $\omega_x$ from spectra like Fig. 2(a), we fit the natural logarithm of the data (to account for linear probe absorption) with the lineshape (\ref{pv}) to determine the spacing between the $v=0$ points of the red and blue SBs.

The dimensionless FWHM of (\ref{pv}) is approximately 4.131.
Using this with (\ref{vFunction}) gives the relation
\begin{align}
\Gamma_\text{SB} \approx 1.033f_x |D| k_B T / U_0
\end{align}
between the FWHM $\Gamma_\text{SB}$ (in temporal frequency units) of the lineshape (\ref{pv}) and the temperature $T$.
Equation (18) then follows from this together with Eq. (4), $|D| = 1$, and rewriting $T = T_\text{SB}$.
Note that for non-magic lattices, the competition of harmonic and anharmonic shifts will lead to both broadening and narrowing effects for SB transitions.


\end{document}